# A Core of E-Commerce Customer Experience based on Conversational Data using Network Text Methodology

Andry Alamsyah [1], Nurlisa Laksmiani [2], Lies Anisa Rahimi [3]
School of Economics and Business, Telkom University, Bandung, Indonesia
[1] andrya@telkomuniversity.ac.id, [2] nurlisalaksmiani@gmail.com, [3] liesanisa87@gmail.com

**Abstract**

*Abstract - E-commerce provide efficient and effective way to exchange goods between sellers and customer. E-commerce has been a popular method for doing business, because its simplicity of having commerce activity transparently available, including costumer voice and opinion about their own experience. Those experiences can be a great benefit to understand customer experience comprehensively, both for sellers and the future customers. This paper applies for e-commerce's and customers in Indonesia.*

*Many Indonesian customer expressed their voice to open social network service such as Twitter and Facebook, where a large proportion of data is in the form of conversational data. By understanding customer behavior through open social network service, we can have the descriptions about the e-commerce services level in Indonesia. Thus, it is related to the government effort to improve Indonesian digital economy ecosystem.*

*A method for finding core topics in large-scale internet unstructured text data is needed, where the method should be fast but sufficiently accurate. Processing a large-scale data is not a straightforward job, it is often need special skills of people and complex software and hardware computer system. We propose a fast methodology of text mining methods based on frequently appeared words and their words association to form network text methodology. This method is adapted from Social Network Analysis by model relationship between words instead of actors.*

*In this paper, we investigate network text performance to analyze customer experience regarding two e-commerce business models, B2C and C2C, where each is represented by two biggest e-commerce sites in Indonesia respectively. Those are Lazada and Mataharimall for B2C, and Bukalapak and Tokopedia for C2C. We identify topics group in network text using Modularity metric. The result is description about each study case customer experience dominant topics that need to be addressed by the company*

## 1. Introduction

The development of information technology, instructing users and companies to use it in accordance with their goals and needs. The increasing penetration of internet users in Indonesia every year is one of the impacts caused by technological developments in business and government. Following the world trend, the popularity of online shopping in Indonesia is also rising. Online shopping is growing as a very attractive trend along with the growth of internet users every year [1]. The wide range of attractive, easy, creative and effective online product and service offerings will also be a factor in accelerating the growth of online shopping transactions also known as e-commerce. Observing the huge potential of online buying and selling ecosystems which are increasingly surrounded by the consumers, no wonder that online shopping activities have now become a lifestyle preferred by many people. In Indonesia, online buying and selling sites, as known as e-commerce, has been develop rapidly. The e-commerce sites in Indonesia can be said to be more dominating to the consumer. The development of e-commerce itself also supported by a number of market size that is constantly increasing from the last report in 2014 [2].

In this research, we investigate *Business-to-Consumer* known as B2C e-commerce type and *Consumer-to-Consumer* known as C2C. Based on *iResearch*, B2C and C2C's market size is constantly rising every year [2]. Therefore, in this research we investigate the popular B2C and C2C e-commerce site in Indonesia; they are *Lazada* and *MatahariMall* as B2C sites and *Tokopedia* and *BukaLapak* as C2C sites.

Similar to offline shopping, customers who buy or visit an e-commerce site, might have their own valuation towards the perceived service. Ease of e-commerce sites can also lead to bad experiences for customers who have made their transactions or just visiting. For every purchase made through the internet, at least there are dissatisfactions involving unclear information, delayed upload sessions for seller, payment problem, delayed delivery, extra charges in the purchase related to delivery, receipt of incomplete orders, and damaged product [3].

Dissatisfied customers who voice their experience on e-commerce site can be a benchmark for the site itself to measure their service level. Based on those phenomena, the government issued a policy about a new e-commerce industry roadmap, which are written in the 14th volume of economic policy package. The details of the policies are, the consumer protection to harmonize the regulations concerning electronic certification, accreditation processes, payment mechanism policies, consumer protection and e-commerce industry performer, and legal action resolution schemes [4].

## 2. Theoretical Background
### 2.1 Customer Experience and E-commerce

Customer experience is a rational and emotional bonding that occurs because of the response to a particular stimulus by optimizing sense (sensory), feel (emotional), think (cognitive), act (action), and relate (relationship) in the marketing efforts before and after the purchase, exchange of information and emotional attachment [5]. Customers learn about a company through their experience gained after making regular purchases and other interactions. Thus, in addition to an increase in customer loyalty [6]. There are also other benefits obtained which are; customers learned more about their own preferences from each company's experience and feedback and the company could learn more about the strengths and weaknesses of each interactions and feedback from their customer experience.

Electronic commerce or e-commerce is the process of delivering information, products, services, and payment processes through telephone lines, internet connections and other digital access. There are seven common types of e-commerce transactions: Business-to-Business (B2B), Business-to-Consumer (B2C), Consumer-to-Business (C2B), Consumer-to-Consumer (C2C), Business-to-Government (B2G), Government-to-Business (G2B) and Government-to-Citizen (G2C). The type of e-commerce that we investigate in this research are B2C, where the seller is a company and the buyer is an individual [7] and C2C, where the consumers selling to other consumers with the help of online market maker [8].

## 2.2    Text Mining

Text mining can be broadly defined as a knowledge-intensive process in which a user interacts with a document collection over time by using a suite of analysis tools. In a manner analogous to data mining, text mining seeks to extract useful information from data sources through the identification and exploration of interesting patterns [9]. One of the used of text mining is to find frequently appeared words from data sources. The result of finding frequently appeared words are a visualize from the contents of a topic and shows a distribution of the vocabulary from each topic mentioned. Moreover, text mining is also used to identify words association, where the words in a sentence would be represented as having an association (relationship) with other words. In this study, the author wanted to know what kind of customer experience gained by customers based on the content of conversations that occur on social media Twitter and Facebook using those methods.

## 2.3    Social Network Analysis

Social Network Analysis (SNA) is a study of the relationship of individuals or other social units, such as an organization, to determine the dependence of the behavior associated with social relationships. In this relationship, described in a node and link. Node is an actor in a network and the link is a line connecting a node with other nodes [10].

We formulate the model as graph *G (N.E)* where *N={n1,n2,...,ni)* is a set of nodes and *E={e1,e2,...,ej}* is a set of edges. |*N*| is a number of nodes in the network and |*E*| is a number of edges in a network. The network has some attributes or certain properties that can be calculated and analyzed. The properties of this network are used to determine the model of a network and analyze it with any other network model called network property. We show several network properties formulation used in this measurement in Table-2.

Modularity metric shows how distinct groups formed in the network. Larger modularity value means clearer boundary between groups in the network. Each group represent certain conversation context, in this paper means certain customer experience topic [11]. The modularity formula is shown below:

$$M = \frac{1}{2m} \sum_{ij} (A_{ij} - \frac{k_i k_j}{2m}) \delta(C_i, C_j)$$

where *M* is modularity value, $A_{ij}$ is number of relations inside group, $\sum k_i k_j/2m$ is the expected number of relations between word *i* and word *j*, $\delta(C_i,C_j)$ is the *kronecker delta coefficient* which equal 1 if *i = j*, and otherwise is 0

## 3. Methodology and Experimental

We use *Text Mining* to find the patterns between text data, including relationship between words. Extracting words relationship into meaning for various purposes [12], we do a three step of pre-processing for deleting irrelevant tweets and comments and produce both tweets and comments which is relevant to the research that is customer experience, followed by finding frequently appeared words to display words that are considered important [13]. We also create association rules between those dominant words to define the relationship between words, with the help of association rules we can find the most common combination of the dominant words that can easily to understand. After this process, we construct network text of dominant word with the use of Social Network Analysis (SNA) method to analyse network text consisting a group of nodes (individual or organization) and edges located between the nodes, therefore we can determine the position and role owned by the actor on the network [14], to analyse the network of issues that appear based on customer experience, we can see how the pattern of interaction occurs in a group of words that are based on modularity or group of word in the network.

The workflow of this research shown in Figure 1.

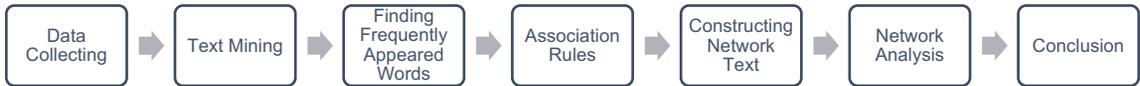

Figure 1. Workflow of this research

The first process is data collection; we directly crawl *Twitter* data using API from software language based and *Facebook* data using web-based software. The data are tweets that come from each e-commerce's Twitter profile and comments in each Facebook sites. After we get the data, we did a three step of pre-processing for deleting irrelevant tweets and comments and produce the data which is relevant to the research that is customer experience. The third process is finding the frequently appeared words by using web-based word cloud generator. The fourth process is to create association rules between dominant words. The fifth process is forming a graph or network text based on the result of association words by using social network analysis method. The sixth process is to analyze the network of customer experience, we can see how the pattern of interaction occurs, the issue or group of words that are based on modularity or group of word in the network. The data profile is shown in Table 1 below.

*Table 1. Data Profile*

|  | Lazada | MatahariMall | Tokopedia | BukaLapak |
|---|---|---|---|---|
| Raw Data | 25.436 data | 3.471 data | 10.518 data | 10.253 data |
| Processed Data | 9.100 data | 3.514 data | 3.692 data | 3.210 data |
| Nodes | 200 | 154 | 65 | 65 |
| Edges | 619 | 648 | 382 | 305 |

The result of association rules for each online business can be seen in Table 2, which is showing the words pair list. Due the limitation space, we only show top 10 of word pair list for each online business which is highly comes up together in a phrase.

*Table 2. Top 10 Word Pairs for Each E-commerce Sites*

| *Lazada Words Pair* | | *MatahariMall Words* | | *Tokopedia Words* | | *BukaLapak Words* | |
|---|---|---|---|---|---|---|---|
| *Words* | *Weight* | *Words* | *Weight* | *Words* | *Weight* | *Words* | *Weight* |
| pesanan-sampai | 116 | pesanan-sampai | 56 | tidak bisa-proses | 100 | dana-belum | 88 |
| pesanan-tidak sampai | 80 | pesanan-tidak sampai | 49 | sampai-sekarang | 86 | sudah-transfer | 71 |
| pesanan-tidak kirim | 75 | pesanan-terima | 45 | sudah-tunggu | 61 | sampai-sekarang | 65 |
| barang-sampai | 62 | tidak respon-admin | 44 | penjual-tidak | 58 | ada-respon | 61 |
| tidak sesuai-estimasi | 55 | tunggu-lama | 38 | resi-invalid | 57 | resi-invalid | 59 |
| barang-tidak sampai | 42 | tunggu-pesanan | 37 | dana-kembali | 54 | transaksi-batal | 59 |
| pesanan-batal | 41 | tunggu-respon | 34 | lama-proses | 54 | barang-belum sampai | 57 |
| batal-sepihak | 36 | komplain-tidak respon | 25 | penjual-penipu | 51 | masih-kosong | 56 |
| pengiriman-lama | 32 | pelayanan-buruk | 19 | ada-respon | 48 | transaksi-status | 51 |
| murah-harga | 30 | pesanan-kirim | 19 | tidak-proses | 45 | aktif-kembali | 51 |

## 4. Result and Analysis

The network text is formed by the words pair using modularity class method. The method is related to its degree centrality and weighted degree for each online business that can be seen in Figure 1. In this network, we can see that each node and edge have different colours and thickness of lines. The difference between those colors indicates groups of words that appear on the network which mean, the words that connected through the network edge have relations with other words in the same topic/issue. The thicker the edge means that those words are often appears together.

The graph of Figure 1.a shows that *Lazada* has denser network text compare others. The overall modularity value is 0.411. There are 11 different colors signify 11 group of words or 11 topics. Their size can also be identified, for example the top-2 biggest group shown by green and blue colour have their consecutive size is 27.84% and 15.98% of overall network size. Issues discussed in green group refers to *"pengiriman"* and *"lama"*, which means the experience felt by customers is the delivery process takes a long time. Issues discussed in the blue network group refers to *"pesanan"* and *"sampai",* which means the customer receives their order.

For network text of *MatahariMall*, we can see in 1.b that its network is similar with *Lazada* which has a high-density network text, with overall modularity value of 0.353. It is similar to Figure 1.a; we can see that Figure 1.b also has eleven different colours in its network text based on their modularity method. It can be seen that 2 groups of networks with the highest issues are shown by the purple and dark blue network with modularity values of 20.13% and 12.99%. The issues discussed in purple network group refers to *"penipuan", "website",* and *"bohong",* which means the customer felt that the *MatahariMall* website is fraudulently perpetrating them. Issues discussed in the dark blue network group refers to the word *"barang", "bagus",* and *"habis",* which means a lot of goods are considered good by the customer, but those goods are often discharged when the customer wants to make a transaction.

Both Figure 1.a and 1.b are showing the network text of online business B2C which are; *Lazada* and *MatahariMall*. As the comparison between online business B2C network, both online businesses have two different issues that shows two different customer experiences. As for *Lazada*, the dominant issue that arises is that the delivery of orders takes a considerable amount of time spent on the *Lazada* website, but there are some who state that orders received even though they (customers) have to wait for a long time. The issue is likely to occur due to several factors. While the dominant issue that arises in *MatahartiMall* is the customer feels that the *MatahariMall* site is often fraudulent, but they also feel that *MatahariMall* sell a good quality product. The summary of issues in both e-commerce sites can be seen in table 3.

Figure 1.c shows a network text of *Tokopedia* that has a high-density network text with the overall modularity value is 0.359. There are 6 different colors that signify 6 group of words or 6 topics mentioned. In this network, their size can also be identified, where top-2 biggest group shown by blue and purple color have their consecutive size is 26.56% and 17.19% of overall network size. Issues discussed in blue group refers to "*transaksi*" while the issues in purple group refers to "*proses*" that is often connected, which is means the experience felt by customers are they could not process a transaction, the transaction is cancelled by the seller, the returns process and refunds were difficult.

For network text of *BukaLapak*, in Figure 1.d, the network is more spread out with the overall modularity value is 0.481. There are 8 different colors signify 8 group of words or 8 topics. Their size can also be identified, in this network, where top-3 biggest group shown by red, blue and green colors have their consecutive size is 23.08% for both red and blue then 15.38% for green of overall network size. Issues discussed in these 3 color groups are "*transaksi*", "*status*" and "*barang*", which means the experience felt by customers such as, the transaction status expired, the transaction status is not updated, the goods do not arrive, the goods do not match and etc.

The last 2 figures, which are Figure 1.c and 1.d, are showing the network text of marketplaces C2C *Tokopedi*a and *BukaLapak*. From the network text shown in Figure 1.c and Figure 1.d, we can assume that both *Tokopedia* and *BukaLapak* have a same issue which associated with their transaction process. As for *Tokopedia* the issue appeared is more directed to their transaction process, while *BukaLapak* is more to the transaction statuses, therefore more issues about complaints form as an experience are received regarding the transaction at *BukaLapak*. The detail of comparison can be seen in table 4. It can also give an overview for both online business that their transaction activity needs a quick and precise action or postured to handle the customer reaction through their experience.

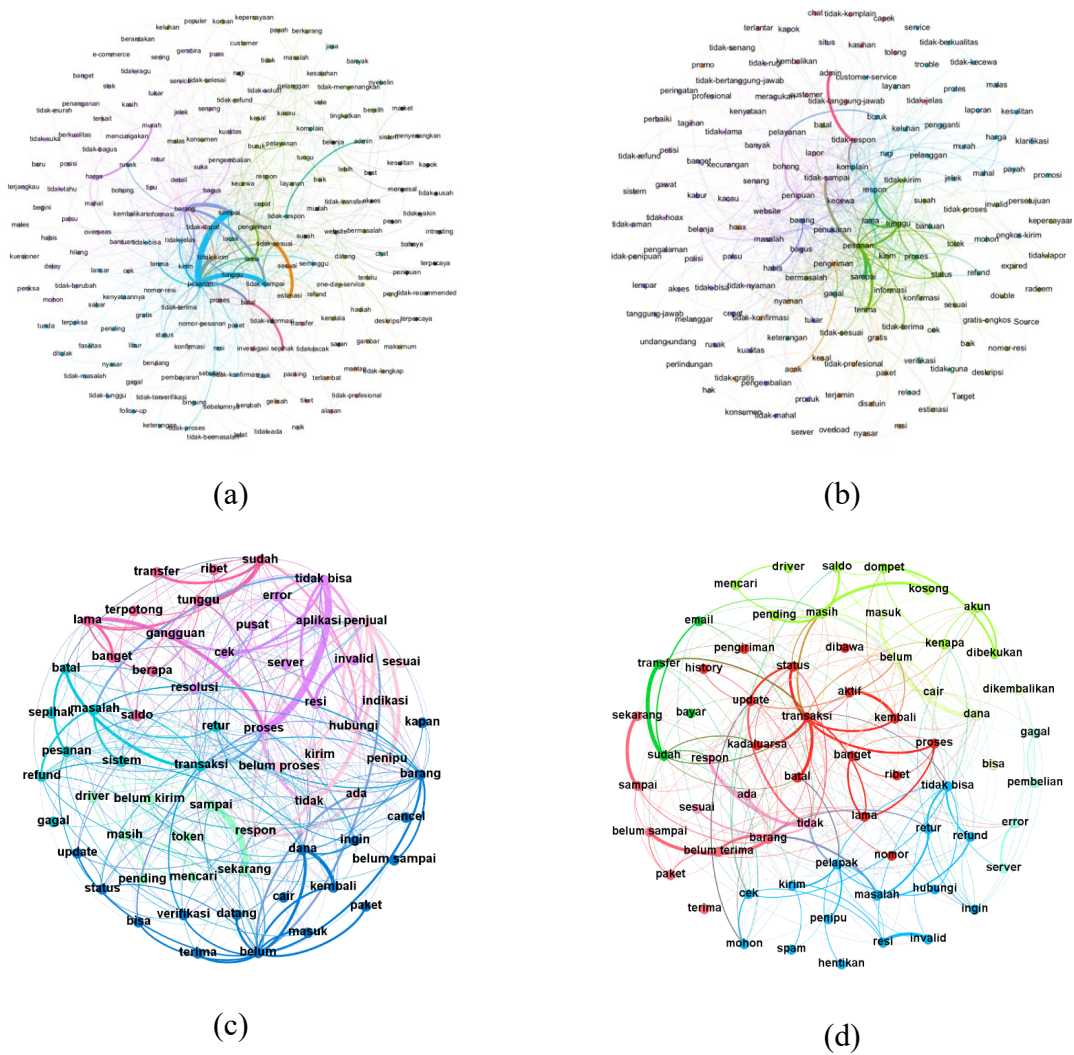

*Figure 1. Network Text of Customer Experience in (a) Lazada, (b) MathariMall, (c) Tokopedia, and (d) BukaLapak*

*Table 3. Issue Comparison on B2C e-Commerce Sites*

| No | Lazada | MatahariMall |
|---|---|---|
| 1. | The two main issues shown based on modularity class are refers to words; **"pengiriman - lama"** and **"pesanan – sampai"**. | The two main issues shown based on modularity class are refers to words; **"penipuan – website - bohong"** and **"barang – bagus - habis"**. |
| 2. | The dominant issue that arises is that **the delivery of orders takes a considerable amount of time** spent on the Lazada website, but there are some who state that orders received even though they (customers) have to wait for a long time. | The dominant issue that arises is the customer feels that **the site is often fraudulent**, but they also feel that MatahariMall **sells a good quality product**. |

*Table 4. Issue Comparison on C2C e-Commerce Sites*

| No | Tokopedia | BukaLapak |
|---|---|---|
| 1. | The issues shown based on modularity class are refers to words **"transaksi"** and **"process"**. | The issues shown based on modularity class are refers to words **"transaksi"**, **"status"**, and **"barang"**. |
| 2. | The main issue is directed to **transaction process**. | The main issue is about the **transaction statuses**. |

In this case of Consumer-to-Consumer (C2C) experience, we figured out that the customer experience is likely about dissatisfied which is turn to be complaints. It can be implied that the transaction between customers is more open to conduct deals through the 3rd parties, they both felt they can act like sellers and buyers so it's easy for them to express their experience directly especially when they felt dissatisfied with the complaints, because they were facing fellow customer not a business to satisfy their needs through the online business experience.

From this analysis, we can determine how customer assessment against the service provided by each online business based on their customer experience form. We can also classify the issues about the service from each online business related to customer experience into network which is more simply to understand.

## 5. Conclusion

As a conclusion, our proposed methodology provides more detail result in summarizing the conversation content on social media and faster to analyze large quantities of documents and unstructured information to extract useful patterns. With the help of network text analysis and social network analysis method we can easily to understand the formulation of graph about the structure of social relationship in the network, and also indicate some issues that appeared in the network by statistical report centrality and modularity.

For general marketing intelligence context, this methodology gives more variation in result based on the network text and also useful for online businesses to determine an action plan to handle their customer. Knowing the experience that felt by customer, whether it is good or bad, are important for online businesses to help them give their customer an optimal service and increase the customer experience. As for experience that leads to a good experience, online business can continue to maintain and even improve its services so that its customers become loyal customers can even help expand the market network of online business, and for some bad experience that were felt, which leads to complaint, those online business can handle it quickly and more appropriate way after learning the contents of customer complaints that being discussed in social media from the result of this research.